\UseRawInputEncoding
\documentclass[aps,onecolumn,10pt]{revtex4}
\usepackage{amsmath}
\usepackage{amssymb}
\usepackage{hyperref}
\hypersetup{colorlinks=true,linkcolor=red,citecolor=green,urlcolor=blue}
\numberwithin{equation}{section}
\numberwithin{equation}{section}

\begin{document}
\allowdisplaybreaks
\setcounter{equation}{0}

\title{Exceptional points for associated Legendre functions of the second kind}

\author{Tianye Liu, Daniel A. Norman and Philip D. Mannheim}
\affiliation{Department of Physics, University of Connecticut, Storrs, CT 06269, USA \\
tianye.liu@uconn.edu,daniel.norman@uconn.edu, philip.mannheim@uconn.edu\\ }

\date{January 8 2023}

\begin{abstract}

We consider the complex $\nu$ plane structure of the associated Legendre  function of the second kind $Q^{-1/2-K}_{\nu}(\cosh\rho)$. We find that for any noninteger value for $K$  $Q^{-1/2-K}_{\nu}(\cosh\rho)$ has an infinite number of poles in the complex $\nu$ plane, but for any negative integer $K$ there are no poles at all. For $K=0$ or any positive integer $K$ there is only a finite number of poles, with there only being one single pole (at $\nu=0$) when $K=0$. This pattern is characteristic of the exceptional points  that appear in a wide variety of physical contexts. However, unusually for theories with exceptional points, $Q^{-1/2-K}_{\nu}(\cosh\rho)$ has an infinite number of them. Other than in the $PT$-symmetry Jordan-block case, exceptional points usually occur at complex values of parameters. While not being Jordan-block exceptional points themselves, the exceptional points associated with the $Q^{-1/2-K}_{\nu}(\cosh\rho)$ nonetheless occur at real values of $K$.

\end{abstract}

\maketitle

\section{Introduction}
\label{S1}

The associated Legendre functions are solutions to the second-order differential equation

\begin{align}
\left[\frac{d^2}{d\rho^2}+\frac{\cosh\rho}{\sinh\rho}\frac{d}{d\rho}-\nu(\nu+1)-\frac{\mu^2}{\sinh^2\rho}\right]F(\nu,\mu,\cosh\rho)=0,
\label{1.1}
\end{align}
and as such they generalize the standard Legendre functions to noninteger $\nu$ and $\mu$. Being second-order differential equations they have two classes of solutions, the first kind being the $P^{\mu}_{\nu}(z)$, the second kind being the $Q^{\mu}_{\nu}(z)$. In general  these solutions can be expressed in terms of hypergeometric functions \cite{AS,GR}: 

\begin{align}
P^{\mu}_{\nu}(\cosh\rho)&=\frac{1}{ \Gamma(1-\mu)}\left(\frac{\cosh\rho+1}{\cosh\rho-1}\right)^{\mu/2}F(-\nu,\nu+1;1-\mu;(1-\cosh\rho)/2),
\nonumber\\
Q^{\mu}_{\nu}(\cosh\rho)&=\frac{e^{i\mu\pi}\pi^{1/2}\Gamma(\nu+\mu+1)}{ 2^{\nu+1}\Gamma(\nu+3/2)}(\sinh\rho)^{\mu}(\cosh\rho)^{-\nu-\mu-1}
F(\nu/2+\mu/2+1,\nu/2+\mu/2+1/2;\nu+3/2;1/\cosh^2\rho).
\label{1.2}
\end{align}
However, for some specific values of $\mu$ and $\nu$ they can be expressed in terms of elementary functions. Thus for $\mu=-1/2$ and arbitrary $\nu$ we have \cite{AS}
\begin{align}
P^{-1/2}_{\nu}(\cosh \rho)&= \left(\frac{1}{2\pi\sinh\rho}\right)^{1/2}\frac{(e^{(\nu+1/2)\rho}-e^{-(\nu+1/2)\rho})}{\nu+1/2},
\nonumber\\
Q^{-1/2}_{\nu}(\cosh \rho)&=- i\left(\frac{\pi}{2\sinh\rho}\right)^{1/2}\frac{e^{-(\nu+1/2)\rho}}{\nu+1/2}.
\label{1.3}
\end{align}
The function $P^{-1/2}_{\nu}(\cosh \rho)$ is well behaved at $\rho=0$, while the function $Q^{-1/2}_{\nu}(\cosh \rho)$ is singular  at $\rho=0$. Similarly, the function $P^{-1/2}_{\nu}(\cosh \rho)$ is well behaved at $\nu=-1/2$, while the function $Q^{-1/2}_{\nu}(\cosh \rho)$ is singular  at $\nu=-1/2$, having a single pole there. An analogous though qualitatively different behavior is met for the general $\mu=-1/2-K$ with arbitrary $K$. Specifically, while $P^{-1/2-K}_{\nu}(\cosh \rho)$ remains singularity free, it is the purpose of this paper to show that for noninteger $K$ $Q^{-1/2-K}_{\nu}(\cosh \rho)$ develops not just one or even two but actually an infinite number of poles in the complex $\nu$ plane.  Moreover, we will also show that for any negative integer $K$ there are no poles at all, while for zero or positive integer $K$ there is only a finite number of poles. Thus for the $Q^{-1/2-K}_{\nu}(\cosh \rho)$ functions the points with integer $K$ form an infinite family of exceptional points. 

Exceptional points are singular points in parameter space where the behavior of a system undergoes a qualitative change \cite{Kato1966}, and such singular points had typically only been known to occur  at complex values of the parameters. However, with the emergence of the $PT$ symmetry program pioneered by Bender and collaborators  both theoretically and experimentally 
\cite{Bender1998,Bender1999,Bender2002,Bender2007,Makris2008,Guo2009,Peng2014,Cummer2016,Mannheim2018a,
ElGanainy2018,Bender2018}, exceptional points have gained  prominence, with the $PT$ setting providing not just a very informative  framework for illustrating some of the general features associated with exceptional points, in the $PT$ case the singularities often occur when parameters are real, to thus make them experimentally accessible. 

The $PT$ program itself is based on the recognition that the physical consistency of a quantum mechanical theory does not require that its Hamiltonian  be Hermitian, despite the fact that with Hermiticity one has both reality of energy eigenvalues and conservation of probability. However, while Hermiticity implies the reality of eigenvalues, there is no converse theorem that says that eigenvalues cannot all be real if the Hamiltonian is not Hermitian, with Hermiticity only being sufficient for reality but not necessary. Similarly, while using the Dirac inner product (the overlap of a ket with its Hermitian conjugate bra) yields probability conservation in the Hermitian Hamiltonian case, there is no theorem that says that one cannot obtain probability conservation with a non-Hermitian Hamiltonian and a different inner product. With the development of the $PT$ program it was realized that the necessary condition for the reality of eigenvalues is that the Hamiltonian have an antilinear symmetry. While any antilinear symmetry would suffice for this purpose, because it was for $PT$ symmetry (parity $P$ is linear and time reversal $T$ is antilinear) that this was first identified, theories of this type are generically referred to  as $PT$ theories. Moreover, in the antilinear symmetry case an inner product that is built out of the overlap of a ket with its antilinear conjugate bra is time independent.

While antilinear symmetry thus reproduces the key features of Hermitian quantum mechanics it actually goes further as it can also achieve things that cannot be obtained in the Hermitian case. In particular if a Hamiltonian $H$ has an antilinear  symmetry and obeys $AH=HA$ where $A$ is an antilinear operator (viz. an operator that not only acts on operators and states but also complex conjugates complex numbers), then its eigenstates obey both $H|\psi\rangle =E|\psi\rangle$ and $AH|\psi\rangle=AE|\psi\rangle=HA|\psi\rangle=E^*A|\psi\rangle$. Thus, as first noted by Wigner in his study of time reversal invariance,  for every eigenvalue $E$ with eigenvector $|\psi\rangle$ there is an eigenvalue $E^*$ with complex conjugate eigenvector $A|\psi\rangle$. One can thus have real eigenvalues ($E=E^*$) or eigenvalues that come in complex conjugate pairs ($E\neq E^*$). This latter option has spurred a large amount of activity especially in the field of experimental optics, as realized through physical systems with (balanced) gain and loss (see e.g. the review of \cite{ElGanainy2018}).

In many $PT$ studies one deals with Hamiltonians in which parameters can be varied continuously, and  one can go from domains in which $E$ and $E^*$ form a complex conjugate pair to domains in which energies are real (the case with $E=E^*$). At the transition point something quite unusual happens, namely not only do the energies become equal ($E_R+iE_I\rightarrow E_R$, $E_R-iE_I\rightarrow E_R$) the complex conjugate eigenvectors become equal too. Thus at the transition point there is a loss of eigenvectors, and the Hamiltonian becomes of nondiagonalizable Jordan-block form with its set of eigenvectors being incomplete. (The emergence of the full set of eigenvectors as we go away from the transition point is referred to as an unfolding, and is studied for the case of a finite number of eigenvectors  in \cite{Graefe2008}.) Away from the transition point and on either side of it the eigenvectors do form complete bases, and the discontinuity at the transition point is in the loss of Hamiltonian eigenvectors, with the lost eigenvectors becoming nonstationary solutions to the Schr\"odinger equation \cite{Bender2008b}. The point in parameter space at which the Hamiltonian becomes Jordan block is thus an exceptional point. It is outside the Hermitian Hamiltonian framework as Hermitian Hamiltonians always have complete sets of eigenvectors. And not only do exceptional points go beyond Hermitian quantum mechanics, systems with exceptional points have now been observed experimentally \cite{ElGanainy2018}, to thus make this option for quantum theory fully viable. 

On the theoretical side these remarks are of relevance to the construction of a quantum-mechanically viable theory of gravity.  In the standard second-order Einstein gravity theory radiative corrections generate fourth-order gravity terms, and it had been thought that such theories have states of negative Dirac inner product and a consequent loss of unitarity. However, it turns out that these theories are not Hermitian theories. Rather, they are $PT$ theories, with the $PT$ theory inner product being time independent, positive definite, and fully unitary \cite{Bender2008a}.

If one now switches off the second-order term, what remains is a pure fourth-order theory of gravity, such  as the conformal gravity theory of interest to us in this paper, with a Hamiltonian that turns out to be Jordan block \cite{Bender2008b}. Thus the limit in which we switch off the second-order term is a singular limit, with the gravity  theory then being at an  exceptional point, while still remaining unitary \cite{Bender2008b}. For conformal gravity the action is of the form \cite{Mannheim2006,Mannheim2017}
\begin{eqnarray}
I_{\rm W}=-\alpha_g\int d^4x\, (-g)^{1/2}C_{\lambda\mu\nu\kappa}
C^{\lambda\mu\nu\kappa},
\label{1.4}
\end{eqnarray}
where 
\begin{eqnarray}
C_{\lambda\mu\nu\kappa}= R_{\lambda\mu\nu\kappa}
-\frac{1}{2}\left(g_{\lambda\nu}R_{\mu\kappa}-
g_{\lambda\kappa}R_{\mu\nu}-
g_{\mu\nu}R_{\lambda\kappa}+
g_{\mu\kappa}R_{\lambda\nu}\right)
+\frac{1}{6}R^{\alpha}_{\phantom{\alpha}\alpha}\left(
g_{\lambda\nu}g_{\mu\kappa}-
g_{\lambda\kappa}g_{\mu\nu}\right)
\label{1.5}
\end{eqnarray}
is the conformal Weyl tensor. The gravitational coupling constant $\alpha_g$ is dimensionless, with the theory thus not only being unitary, it is also power-counting renormalizable. Through its exceptional point structure conformal gravity is thus offered as a fully consistent theory of quantum gravity. 

When the conformal gravity theory is applied to cosmology the three-curvature of the associated  Robertson-Walker background geometry is found to be negative, i.e., topologically open \cite{Mannheim2006,Mannheim2017}. Moreover,  fluctuations around this background are  found to obey none other than the associated Legendre function equation given in (\ref{1.1}) with $\rho$ being the conformal time  \cite{Mannheim2020,Amarasinghe2021b}.   (Technically one finds that the relevant $\nu$ are of the conical function form $\nu=-1/2+i\tau$ with a real $\tau$ that lies in the range $0\leq \tau \leq \infty$, so one can consider analyticity in $\nu$ or in $\tau$.) As we now show, the conformal gravity  fluctuation equations have an exceptional  point structure. Thus conformal gravity actually has two exceptional point structures associated with it, one for the background and the other for the fluctuations around it. While it was conformal gravity that led the current authors to (\ref{1.1}), the reader can regard our paper purely as a study of modes propagating in a space of constant negative curvature.  What makes the $Q^{-1/2-K}_{\nu}(\cosh\rho)$ so interesting is that  being at an exceptional point typically involves the loss of a finite number of dynamical functions. However,  for the $Q^{-1/2-K}_{\nu}(\cosh\rho)$ with integer $K$ we lose an infinite number. Moreover, with every integer value of $K$ itself being an exceptional point, in total we even have an infinite number of exceptional points. And not only that, even though the exceptional points associated with the $Q^{-1/2-K}_{\nu}(\cosh\rho)$ are not themselves of the Jordan-block type, they are nonetheless real, even though non-Jordan-block exceptional points typically are not.  Associated Legendre functions of the second kind thus provide an interesting  theoretical  laboratory for exploring some of the general features of exceptional points. 

\section{Complex $\nu$ plane  structure of associated Legendre functions functions }
\label{S2}
There are various ways to explore the analytic structure of the $Q^{-1/2-K}_{\nu}(\cosh\rho)$ in the complex $\nu$ plane. The most direct is to note that since the location of the $\nu$ plane poles does not depend on the value of $\cosh\rho$, we can study  $Q^{-1/2-K}_{\nu}(\cosh\rho)$ at large $\cosh\rho$ since then it can be expressed in terms of elementary functions, viz. \cite{GR}
\begin{align}
Q^{-1/2-K}_{\nu}(\cosh\rho) \rightarrow  -\frac{i\pi^{1/2} e^{-i\pi K}\Gamma(\nu+1/2-K)}{2^{\nu+1}\Gamma(\nu+3/2)}(\cosh\rho)^{-\nu-1}.
\label{2.1}
\end{align}
$Q^{-1/2-K}_{\nu}(\cosh\rho)$ thus possesses poles at $\nu=K-1/2, K-3/2, K-5/2,...$ and zeroes at $\nu=-3/2,-5/2,-7/2,...$. Since the positions of the zeroes  are independent of $K$, for any noninteger $K$  $Q^{-1/2-K}_{\nu}(\cosh\rho)$ has an infinite number of complex $\nu$ plane poles. However, when $K$ is an integer some of the poles can be cancelled, with all of them being cancelled when $K$ is a negative integer. When $K=0$ one pole remains (at $\nu=-1/2$, just as shown in (\ref{1.3})), when $K=1$ two poles remain (at $\nu=-1/2,-3/2$), and so on.  The full pattern is 
\begin{align}
& K=-2~~{\rm no~poles}, \qquad -2<K<-1~~{\rm infinite~set~of~poles},
\nonumber\\
& K=-1~~{\rm no~poles}, \qquad -1<K<0~~{\rm infinite~set~of~poles},
\nonumber\\
& K=0~~{\rm one~pole}, \qquad 0<K<1~~{\rm infinite~set~of~poles},
\nonumber\\
& K=1~~{\rm two~poles}, \qquad 1<K<2~~{\rm infinite~set~of~poles},
\nonumber\\
& K=2~~{\rm three~poles}, \qquad 2<K<3~~{\rm infinite~set~of~poles}.
\label{2.2}
\end{align}
We thus obtain an infinite set of exceptional points,  one for each integer value of $K$.

An alternate way to explore the complex $\nu$ plane structure of the $Q^{-1/2-K}_{\nu}(\cosh\rho)$ is to first relate the $Q^{-1/2-K}_{\nu}(\cosh\rho)$ to the $P^{-\nu-1/2}_{K}(\coth\rho)$ through the so-called Whipple relation \cite{AS,GR}
\begin{align}
Q^{-1/2-K}_{\nu}(\cosh\rho)
=-ie^{-iK\pi}\left(\frac{\pi}{2\sinh\rho}\right)^{1/2}\Gamma(\nu+1/2-K)P^{-\nu-1/2}_{K}(\coth\rho).
\label{2.3}
\end{align}
And then on showing that the $P^{-\nu-1/2}_{K}(\coth\rho)$ have no complex $\nu$ plane poles,  we can then identify the poles of $Q^{-1/2-K}_{\nu}(\coth\rho)$ as the poles of $\Gamma(\nu+1/2-K)$, unless they are cancelled by any relevant zeroes that the $P^{-\nu-1/2}_{K}(\coth\rho)$ might and in fact do have.

To explore the complex $\nu$ plane structure of the $P^{-\nu-1/2}_{K}(\coth\rho)$ we use their representation in terms of hypergeometric functions. For the hypergeometric representation we have to convert $\coth\rho$ to $\cosh\rho$ in order to use (\ref{1.2}). To this end we set $\coth\rho=\cosh\alpha$, so that $\sinh\rho=1/\sinh\alpha$ and $\cosh\rho=\coth\alpha$. Since (\ref{1.2}) would hold for $\cosh\alpha$, for $\coth\rho$ we obtain 
\begin{align}
&P^{-\nu-1/2}_{K}(\coth\rho)=\frac{1}{\Gamma(\nu+3/2)}e^{ -(\nu+1/2)\rho}F(-K,K+1;\nu+3/2;(1-\coth \rho)/2)
\nonumber\\
&=\frac{1}{\Gamma(\nu+3/2)}e^{ -(\nu+1/2)\rho}\left(  1-\frac{K(K+1)(1-\coth\rho)}{2(\nu+3/2)}-\frac{K(-K+1)(K+1)(K+2)(1-\coth\rho)^2}{8(\nu+3/2)(\nu+5/2)}....\right)
\nonumber\\
&=e^{ -(\nu+1/2)\rho}\left( \frac{1}{\Gamma(\nu+3/2)}-\frac{K(K+1)(1-\coth\rho)}{2\Gamma(\nu+5/2)}-\frac{K(-K+1)(K+1)(K+2)(1-\coth\rho)^2}{8\Gamma(\nu+7/2)}....\right).
\label{2.4}
\end{align}
Thus no individual term in $P^{-\nu-1/2}_{K}(\coth\rho)$ has any poles in $\nu$. However, $P^{-\nu-1/2}_{K}(\coth\rho)$ does have zeroes due to the presence of  the $1/\Gamma(\nu+3/2)$  factor. We thus obtain exactly the same $\Gamma(\nu+1/2-K)/\Gamma(\nu+3/2)$ patten for the $Q^{-1/2-K}_{\nu}(\cosh\rho)$ as obtained from the large $\cosh\rho$ expansion, just as we should.

While the sum in (\ref{2.4}) only has a finite number of terms in it if $K$ is an integer, it is still possible that with noninteger $K$ the then infinite sum in (\ref{2.4}) might diverge even though no individual term in it does. However, this is not the case since through the 
$\Gamma(\nu+1/2-K)/\Gamma(\nu+3/2)$ factor that appears in the large $\cosh\rho$ limit all the poles have already been accounted for, and thus no further poles can be generated from the hypergeometric expansion approach that have not already been obtained from the selfsame $\Gamma(\nu+1/2-K)/\Gamma(\nu+3/2)$ factor that it equally possesses.	

From analysis of the last line in (\ref{2.4}) we can understand the pattern we have found. The $1/\Gamma(\nu+3/2)$ term generates zeros at $\nu=-3/2,-5/2,-7/2,...$.  The $1/\Gamma(\nu+5/2)$ term generates zeros at $\nu=-5/2,-7/2,...$. The $1/\Gamma(\nu+7/2)$ term generates zeros at $\nu=-7/2,...$. If $K=0$ all terms in the last line in (\ref{2.4}) other than the  $1/\Gamma(\nu+3/2)$ term are cancelled, the zeroes begin at $\nu=-3/2$, and  with  $\Gamma(\nu+1/2-K)\rightarrow \Gamma(\nu+1/2)$  there is just one pole (at $\nu=-1/2$). If $K=1$ all terms other than the  $1/\Gamma(\nu+3/2)$ and  $1/\Gamma(\nu+5/2)$ are cancelled, so now the zeroes begin at $\nu=-5/2$ ($1/\Gamma(\nu+5/2)$ does not vanish at $\nu=-3/2$), and  with  $\Gamma(\nu+1/2-K)\rightarrow \Gamma(\nu-1/2)$  there are now two poles (at $\nu=-1/2$, $\nu=-3/2$). This pattern repeats for $K=3$ and so on, with each increase in $K$ by one leading to one more pole \cite{footnote1}. If $K=-1$ only the $1/\Gamma(\nu+3/2)$ term is not cancelled and zeros at $\nu=-3/2,-5/2,-7/2,...$ are generated. At the same time $\Gamma(\nu+1/2-K)\rightarrow \Gamma(\nu+3/2)$ and so all poles are cancelled. This then repeats for $K=-2$, $K=-3$ and so on with all poles being cancelled. If $K$ is not an integer then none of the poles in $\Gamma(\nu+1/2-K)$ can be cancelled and there is an infinite number of poles.  To understand how we are able to go from an infinite number of poles if $K$ is not an integer to a finite number if $K$ is an integer we consider $K=0$. As long as $K$ is not zero but is very close to zero, every term in the last line in (\ref{2.4}) is nonzero and there is no cancellation of any pole. When we let $K$ go to zero every term in  this last line is cancelled except the very first one, and all but one of the poles are cancelled. It is in this way that the integer $K$ become exceptional points.

\section{Contour integrals involving $Q^{-1/2-K}_{-1/2+i\tau}(\cosh\rho)$}
\label{S3}

On setting $\nu=-1/2+i\tau$ we can rewrite (\ref{1.1}) in the form 
\begin{align}
\left[\frac{d^2}{d\rho^2}+\frac{\cosh\rho}{\sinh\rho}\frac{d}{d\rho}+\frac{1}{4}-\frac{(-1/2-K)^2}{\sinh^2\rho}\right]Q^{-1/2-K}_{-1/2+i\tau}(\cosh\rho)=-
\tau^2Q^{-1/2-K}_{-1/2+i\tau}(\cosh\rho),
\label{3.1}
\end{align}
and treat it as an eigenvalue equation for eigenfunction $Q^{-1/2-K}_{-1/2+i\tau}(\cosh\rho)$ and eigenvalue $-\tau^2$, where $0\leq \tau \leq \infty$. From (\ref{2.3}) we obtain 
\begin{align}
Q^{-1/2-K}_{-1/2+i\tau}(\cosh\rho)
=-ie^{-iK\pi}\left(\frac{\pi}{2\sinh\rho}\right)^{1/2}\Gamma(i\tau-K)P^{-i\tau}_{K}(\coth\rho),
\label{3.2}
\end{align}
and then from (\ref{3.2}) it follows that
\begin{align}
Q^{-1/2-K}_{-1/2}(\cosh\rho)
=-ie^{-iK\pi}\left(\frac{\pi}{2\sinh\rho}\right)^{1/2}\Gamma(-K)P^{0}_{K}(\coth\rho),
\label{3.3}
\end{align}
with $Q^{-1/2-K}_{-1/2}(\cosh\rho)$ becoming singular at $K=0,1,2,...$.
For the  $Q^{-1/2-K}_{-1/2+i\tau}(\cosh\rho)$ modes the normalization is given by 
\begin{align}
I(K,\rho)=\int_{0}^{\infty}d\tau Q^{-1/2-K}_{-1/2+i\tau}(\cosh\rho)[Q^{-1/2-K}_{-1/2+i\tau}(\cosh\rho)]^*.
\label{3.4}
\end{align}
As we now show, an analysis of (\ref{3.4}) that will dovetail with the singularity in $Q^{-1/2-K}_{-1/2}(\cosh\rho)$  at $K=0$ will provide us with further insight into the nature of exceptional point at  $K=0$, where (\ref{3.2}) has a pole at $\tau=0$. 

To this end use of (\ref{3.2}) enables us to rewrite the integrand  in (\ref{3.4}) as 
\begin{align}
\left[Q^{-1/2-K}_{-1/2+i\tau}(\cosh\rho)\right]\left[Q^{-1/2-K}_{-1/2+i\tau}(\cosh\rho)\right]^*
&=\frac{\pi}{2\sinh\rho}\Gamma(i\tau-K)\Gamma(-i\tau-K)P^{-i\tau}_K(\coth\rho)P^{i\tau}_K(\coth\rho),
\label{3.5}
\end{align}
so that
\begin{align}
I(K,\rho)&=\frac{\pi}{2\sinh\rho}\int_{0}^{\infty}d\tau \Gamma(i\tau-K)\Gamma(-i\tau-K)P^{-i\tau}_K(\coth\rho)P^{i\tau}_K(\coth\rho)
\nonumber\\
&=\frac{\pi}{4\sinh\rho}\int_{-\infty}^{\infty}d\tau \Gamma(i\tau-K)\Gamma(-i\tau-K)P^{-i\tau}_K(\coth\rho)P^{i\tau}_K(\coth\rho).
\label{3.6}
\end{align}
For large $\tau$ the integrand behaves as \cite{Cohl2018}
\begin{align}
\left[Q^{-1/2-K}_{-1/2+i\tau}(\cosh\chi)\right]\left[Q^{-1/2-K}_{-1/2+i\tau}(\cosh\chi)\right]^*
\rightarrow \frac{\pi}{2 \sinh\chi}\tau^{-2-2K},
\label{3.7}
\end{align}
and thus the integral will exist if $-1-2K<0$, i.e $-1/2<K$. For any general $K$ that satisfies this condition, we can integrate $I(K,\rho)$ as an integral on the real $\tau$ axis,  though it  can only be done numerically. However, the integral can also be done as a contour integral, and this will enable us to monitor how the infinite number of poles collapse into a single one when $K=0$.

With the asymptotic bound in (\ref{3.7}) holding for $|\tau |$, it holds on the circle at infinity in the complex $\tau$ plane. The circle contribution to a contour integration will thus be suppressed if $-1<K$, a constraint that holds for our previous $-1/2<K$ condition.
 For $-1/2<K$ we can close the contour in either the upper- or lower-half complex $\tau$ planes.  For definitiveness we shall close above. 
For $-1/2<K<0$ the poles in $\Gamma(i\tau-K)$ are in the upper-half plane while those  in $\Gamma(-i\tau-K)$ are in the lower-half plane. 
The poles in $\Gamma(i\tau-K)$ are on the imaginary $\tau$ axis, being of the form $\tau=i(n-K)$, where $n=0,1,2,...$. Closing the contour in the upper-half plane  then gives 
\begin{align}
I(K,\rho)=\frac{\pi^2i}{2\sinh\rho}\sum_{n=0}^{n=\infty} \frac{(-1)^n}{in!}\Gamma(n-2K)P^{K-n}_K(\coth\rho)P^{n-K}_K(\coth\rho).
\label{3.8}
\end{align}
In order to set $K=0$ we need to treat the ensuing pole at $i\tau=0$. Since $K$ had been taken to be negative, $K$ approaches $K=0$ from below. We thus set $K=-\epsilon$ where $\epsilon$ is positive. Thus at $K=0$ we set 
\begin{align}
I(0,\rho)=\frac{\pi^2}{2\sinh\rho}\sum_{n=0}^{n=\infty} \frac{(-1)^n}{n!}\Gamma(n+2\epsilon)P^{-n}_0(\coth\rho)P^{n}_0(\coth\rho).
\label{3.9}
\end{align}
To evaluate this expression we need to determine $P^{-n}_0(\coth\rho)$. Recalling the general relation \cite{AS,GR}
\begin{align}
P^{-\mu}_{\nu}(z)=\frac{\Gamma(\nu-\mu+1)}{\Gamma(\nu+\mu+1)}\left[P^{\mu}_{\nu}(z)-\frac{2}{\pi}e^{-i\mu\pi}\sin(\mu\pi)Q^{\mu}_{\nu}(z)\right],
\label{3.10}
\end{align}
we obtain
\begin{align}
P^{-n}_{0}(z)=\frac{\Gamma(-n+1)}{\Gamma(n+1)}\left[P^{n}_{0}(z)-\frac{2}{\pi}e^{-in\pi}\sin(n\pi)Q^{n}_{0}(z)\right].
\label{3.11}
\end{align}
We can eliminate the $\sin(n\pi)Q^{n}_{0}(z)$ term since $Q^{n}_{0}(z)$ is not singular. Specifically, we can determine $Q^{n}_{0}(z)$
using the relations 
\begin{align}
Q_{0}(z)=\frac{1}{2}\ln\left(\frac{z+1}{z-1}\right),\quad Q^{n}_{0}(z)=(z^2-1)^{n/2}\frac{d^n}{dz^n}Q_{0}(z),
\label{3.12}
\end{align}
and thus confirm that $Q^{n}_{\nu}(z)$ does not develop a complex $\nu$ plane singularity as $\nu\rightarrow 0$. Consequently, (\ref{3.11}) and (\ref{3.9}) respectively reduce to 
\begin{align}
P^{-n}_{0}(z)=\frac{\Gamma(-n+1)}{\Gamma(n+1)}P^{n}_{0}(z),
\label{3.13}
\end{align}
\begin{align}
I(0,\rho)=\frac{\pi^2}{2\sinh\rho}\sum_{n=0}^{n=\infty} \frac{(-1)^n}{n!}\Gamma(n+2\epsilon)\frac{\Gamma(-n+1)}{\Gamma(n+1)}[P^{n}_0(\coth\rho)]^2.
\label{3.14}
\end{align}
Now $\Gamma(-n+1)$ has single poles at $n=1$, $n=2$, .... However the $P^n_0(\cosh\rho)$ with integer $n$ are ordinary Legendre polynomials, and they are zero if $n>0$. Then since the $P^n_0(\cosh\rho)$ term appears in a squared term, all contributions with $n>0$ are cancelled and all that is left is the $n=0$ contribution
\begin{align}
I(0,\rho)=\frac{\pi^2}{2\sinh\rho}\Gamma(2\epsilon)[P^{0}_0(\coth\rho)]^2=\frac{\pi^2}{2\sinh\rho}\frac{1}{2\epsilon}.
\label{3.15}
\end{align}
Thus the infinite tower of poles with $\tau=i(n-K)$ collapses into just a single $n=0$ pole contribution at $\tau=0$ when $K=0$. To confirm that $I(0,\rho)$ is singular we write it out explicitly using the exact form given in (\ref{1.3}), viz. the $\epsilon \rightarrow 0$ limit  
\begin{align}
I(0,\rho)=\frac{\pi}{2\sinh\rho}\int _0^{\infty}\frac{d\tau}{\tau^2}\rightarrow \frac{\pi}{2\sinh\rho}\int _0^{\infty}\frac{d\tau}{\tau^2+\epsilon^2}
=\frac{\pi}{2\sinh\rho}\frac{\pi}{2\epsilon}= \frac{\pi^2}{4\sinh\rho}\frac{1}{\epsilon},
\label{3.16}
\end{align}
to thus diverge at $\tau=0$. To compare directly with (\ref{3.15}) we evaluate the integral in (\ref{3.16}) as a contour integral. While (\ref{1.3}) is the $K=0$ limit of (\ref{2.3}), since $\Gamma(i\tau-K)\Gamma(-i\tau-K)$ contains poles in both the upper- and lower-half planes we must take one of the $\tau=0$ poles in $I(0,\rho)$ to lie in the upper-half plane and the other in the lower-half plane. Thus closing above gives 
\begin{align}
I(0,\rho)&=\frac{\pi}{4\sinh\rho}\int _{-\infty}^{\infty}\frac{d\tau}{(\tau^2+\epsilon^2)}=\frac{\pi}{4\sinh\rho}\int _{-\infty}^{\infty}\frac{d\tau}{(\tau-i\epsilon)(\tau+i\epsilon)}
=\frac{2i\pi^2}{4\sinh\rho}\frac{1}{2i\epsilon}=\frac{\pi^2}{4\sinh\rho}\frac{1}{\epsilon},
\label{3.17}
\end{align}
which we  recognize as (\ref{3.15}). Thus even for the normalization integral  an infinite number of poles collapses into a single one at the $K=0$ exceptional point. 

\begin{acknowledgments}
One of us (PDM)  wishes to acknowledge helpful discussions with Dr. U. G\"unther.
\end{acknowledgments}


\begin{thebibliography}{99}

\bibitem{AS} M. Abramowitz and I. A. Stegun, \textit{Handbook of Mathematical Functions},  Dover Publications, New York (1972).

\bibitem{GR}  I. S. Gradshteyn and I. M. Ryzhik,\textit{Tables of Integrals, Series, and Products}, Academic Press, New York (1980). 

\bibitem{Kato1966} T. Kato, \textit{Perturbation theory for linear operators},  Springer-Verlag, Berlin (1966).


\bibitem{Bender1998}C. M. Bender and S. Boettcher,  \href{https://doi.org/10.1103/PhysRevLett.80.5243}{Phys. Rev. Lett. \textbf{80}, 5243 (1998).}

\bibitem{Bender1999} C. M. Bender, S. Boettcher and P. N. Meisinger, \href{https://doi.org/10.1063/1.532860}{J. Math. Phys. \textbf{40}, 2201 (1999).}

\bibitem{Bender2002} C. M.Bender, D. C. Brody and H. F. Jones, \href{https://doi.org/10.1103/PhysRevLett.89.270401}{Phys. Rev. Lett. \textbf{89}, 270401 (2002).}

\bibitem{Bender2007} C. M. Bender, \href{https://doi.org/10.1088/0034-4885/70/6/R03}{Rep. Prog. Phys.  \textbf{70}, 947 (2007).}

\bibitem{Makris2008} K. G. Makris, R. El-Ganainy, D. N. Christodoulides, and Z. H. Musslimani, \href{https://doi.org/10.1103/PhysRevLett.100.103904}{Phys. Rev. Lett.  \textbf{100}, 103904 (2008).}

\bibitem{Guo2009} A. Guo, G. J. Salamo, D. Duchesne, R. Morandotti, M. Volatier-Ravat,
V. Aimez, G. A. Siviloglou and D. N. Christodoulides, \href{https://doi.org/10.1103/PhysRevLett.103.093902}{Phys. Rev. Lett. \textbf{103}, 093902 (2009).} 


\bibitem{Peng2014}  B. Peng, S. K. \"Ozdemir, F. Lei, F. Monifi, M. Gianfreda, G. L. Long, S. Fan,
F. Nori, C. M. Bender and L. Yang, \href{https://doi.org/10.1038/nphys2927}{Nature Physics  \textbf{10}, 394 (2014).}

\bibitem{Cummer2016} S. A. Cummer, J. Christensen and  A. Al\`u, \href{https://doi.org/10.1038/natrevmats.2016.1}{
Nature Reviews Material \textbf{1}, 16001 (2016).} 

\bibitem{Mannheim2018a}  P. D. Mannheim, \href{https://doi.org/10.1088/1751-8121/aac035}{J. Phys. A: Math. Theor. \textbf{51}, 315302 (2018).}


\bibitem{ElGanainy2018} R. El-Ganainy, K. G. Makris, M. Khajavikhan, Z. H. Musslimani, S. Rotter and D. N. Christodoulides, \href{https://doi.org/10.1038/nphys4323}{Nature Phys. \textbf{14}, 11 (2018).}

\bibitem{Bender2018} C. M. Bender, \textit{ PT Symmetry in Quantum And Classical Physics}, World Scientific Press, Singapore (2018).

\bibitem{Graefe2008} E. M. Graefe, U. G\"unther, H. J. Korsch and A. E. Niederle, \href{https://doi.org/10.1088/1751-8113/41/25/255206}{J. Phys. A: Math. Theor. \textbf{41}, 255206 (2008).} 



\bibitem{Bender2008b} C. M. Bender and P. D. Mannheim, \href{https://doi.org/10.1103/PhysRevD.78.025022}{Phys. Rev. D  \textbf{78}, 025022 (2008).}

\bibitem{Bender2008a} C. M. Bender and P. D. Mannheim, \href{https://doi.org/10.1103/PhysRevLett.100.110402}{Phys. Rev. Lett. \textbf{100}, 110402 (2008).}

\bibitem{Mannheim2006} P. D. Mannheim,  \href{https://doi.org/10.1016/j.ppnp.2005.08.001}{Prog. Part. Nucl. Phys. \textbf{56}, 340 (2006).}

\bibitem{Mannheim2017} P. D. Mannheim,  \href{http://dx.doi.org/10.1016/j.ppnp.2017.02.001}{Prog. Part. Nucl. Phys. \textbf{94}, 125 (2017).}



\bibitem{Mannheim2020} P. D. Mannheim, \href{https://doi.org/10.1103/PhysRevD.102.123535} {Phys. Rev. D \textbf{102}, 123535 (2020).}

\bibitem{Amarasinghe2021b} A. Amarasinghe, T. Liu, D. A. Norman and  P. D. Mannheim, \href{https://doi.org/10.1103/PhysRevD.103.104022}{Phys. Rev. D \textbf{103}, 104022 (2021).}

\bibitem{footnote1} If both $-\nu-1/2$ and $K$ are nonnegative integers the $P^{-\nu-1/2}_{K}$ become ordinary Legendre polynomials $P^m_{\ell}$, and they vanish if $m>\ell$.

\bibitem{Cohl2018} H. S. Cohl, T. H. Dang and T. M. Dunster, \href{https://doi.org/10.3842/SIGMA.2018.136}{SIGMA \textbf{14}, 136 (2018).}

\end{thebibliography}
\end{document}